# DETERMINATION OF SIZE OF THE EMITTING REGION IN ECLIPSING CATACLYSMIC VARIABLE STARS

I.L. Andronov [1], K.D. Andrych [2]

[1] Department "High and Applied Mathematics, Odessa National Maritime University, Mechnikova Str., 34, 65029, Odessa, Ukraine, *tt_ari@ukr.net*

[2] Department of Astronomy, Odessa National University, Shevchenko Park, 65014, Odessa, Ukraine

ABSTRACT. The dependencies of the phase of eclipse of the white dwarf's centre and the durations of the ascending and descending branches of the light curve on the binary system's parameters were computed using the spherically-symmetric approximation and the more accurate model of the elliptical projection onto the celestial sphere of the companion (red dwarf) that fills its Roche lobe. The parameters of eclipses in the classical eclipsing polar OTJ 071126+440405 = CSS 081231:071126+440405 were estimated.

**Keywords:** Stars: variable – stars: binary – stars: cataclysmic

## 1. Introduction

Interacting binary stars are natural laboratories with extreme conditions that are inaccessible in terrestrial laboratories. There are many different processes that allow to determine physical characteristics of binary systems using methods for mathematical modeling of observations.

Especially useful are studies of eclipsing variables, particularly, eclipsing cataclysmic variables, which allow to determine some physical parameters from photometric observations only.

On the New Year night 31.12.2008, Denisenko and Korotkiy (2009) discovered the unique eclipsing polar (OTJ 071126 + 440405) in the Camelopardalis constellation. The light curve shows that the system has a very short duration of the ascending and descending branches of the primary minimum, what indicates a very small size of the radiation source.

Our theoretical work has been done for interpreting the observational data.

## 2. Effective Dimensions of the Stars

The simplest classical approach is to apply a model of spherical stars not only in Algol-type systems (Shulberg, 1971; Tsessevich, 1980; Chinarova, 2006), but also in cataclysmic systems (Shafter A., 1984; Horne, 1985; Downes et al., 1986; Garnavich et al., 1990 and, more recently, Aungwerojwit et al., 2012), assuming that the red dwarf (which fills its Roche lobe and thus is tidally distorted) is spherical.

Typically, the radius of this sphere $R_2$ is defined as

$$R_2 = \left(\frac{3V}{4\pi}\right)^{1/3},$$

where $V$ is the volume of the Roche lobe, which may be estimated from a suitable approximation of Eggleton (1983)

$$\frac{R_2}{A} = \frac{Cq^{2/3}}{Dq^{2/3} + \ln(1+q^{1/3})},$$

where $q=M_2/M_1$ is the mass ratio, $M_1$ – the mass of the compact primary, $M_2$ – the mass of the secondary which fills its Roche lobe. The values of the coefficients were adopted to be $C$=0.49 and $D$=0.6. The asymptotical approximations for this formula are $R_2/A=Cq^{1/3}$ for $q \ll 1$ and $R_2/A=C/D$ for $q \gg 1$. This expression is a good approximation also for intermediate values of $q$.

However, the Roche lobe is definitely not spherical, having the largest size along the line of centers, and the smallest in the "polar" direction (along the rotational axis).

For our task – modeling of entry/exit time (duration of the descending/ascending branch of the light curve) at the eclipse as functions of the parameters of the binary system (including the size of the white dwarf). We have chosen two models: the popular model of "spherical" red dwarf, and much more accurate model of it's elliptical projection onto the celestial sphere (there is assumed that red dwarf filed it's Roche lobe)

## 3. Calculating the Size of the Emitting Area

Let us consider a cataclysmic binary system, in which red and white dwarf obscure each other. Because these systems are very close (have a small orbital period), they have circular orbits, and this facts greatly simplifies the calculation. For convenience, we consider a coordinate system with its center in the red dwarf, so the white dwarf rotates around it.

The center of the white dwarf during its orbital motion shows an ellipse. The coordinates $x$ and $y$ can be expressed as:

$$x = A \cdot \sin 2\pi\varphi, \quad y = B \cdot \cos 2\pi\varphi, \quad B = A \cdot \cos i,$$

where $\varphi$ is the phase expressed in units of the orbital period $P$, thus a multiplier of $2\pi$ is needed for conversion to radians. It rises to unity that mean the system made full rotation.

Here $A$ and $B$ correspond to minor and major axis of ellipse, which is the projection of the circular orbit of a radius $A$ on the picture area (an area that passes through the center of the first stars perpendicularly to the line of sight). Inclination of the orbit $i$ is the angle between the line of sight and the rotation axis of the binary system.

The distance between the centers of the stars can be calculated by the Pythagorean theorem. Taking into account the above formulas, it may be written in the following form:

$$R^2 = A^2 (1 - \cos^2 2\pi\varphi \cdot \sin^2 i).$$

After some transformations, we get the following form of the same equation, which is convenient to determine the phase corresponding to the projected distance $R$:

$$2\pi\varphi = \arccos\left(\frac{\sqrt{1-\frac{R^2}{A^2}}}{\sin i}\right)$$

We consider that, at the time of internal contact, the distance between the centers equals to the difference in the stellar radii, and at point of external contact – to sum of their radii. Obviously, for convenience, one can use the relative values of the stellar radii.

The phases of contacts were calculated as well as the phase difference between external and internal contacts, which corresponds to the length of ascending (or descending) branch of the eclipse and is measured in units of the orbital period. Results are presented in Fig.1 and Fig.2, where the radius of the white dwarf and the inclination of the orbit were fixed, respectively.

As in cataclysmic binary systems one of the stars – the red dwarf – fills its Roche lobe, it is deformed. To within a few hundredths of a percent profile stars can be approximated to an ellipse (Andronov, 1992).

The limb of the red dwarf indicated in Fig. 3 by a red line. The ellipse – like curves $N_1$ and $N_2$ are the locuses of points distant from the limb of the red dwarf by the radius of the white dwarf. Coordinates of these points can be calculated by these formulas, respectively.

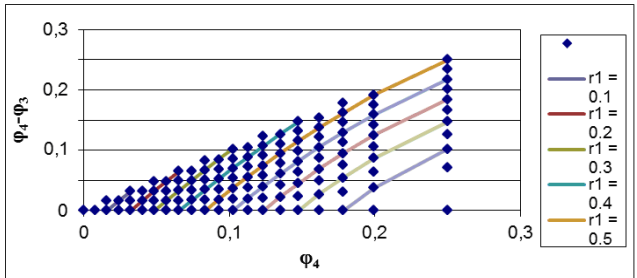

Figure 1: Phase of the external contact and the duration of the ascending branch of the light curve for different pairs of values of the relative radii $r_1$ and $r_2$, varying in increments of 0.05 in the range $0 \leq r_2 \leq r_1 \leq (1-r_2)$. Inclination $i = 90°$.

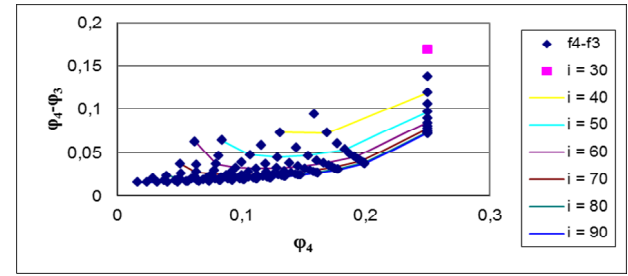

Figure 2: Phase of the external contact and the duration of the ascending branch of the light curve for different pairs of values of the relative radius $r_1$, which varies in increments of 0.05 in the range $0 \leq r_2 \leq r_1 \leq (1-r_2)$, and inclination of the orbit from $30°$ to $90°$ in steps of $5°$. The parameter $r_2=0.05$.

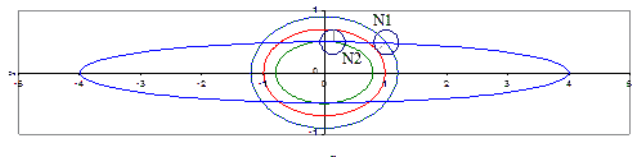

Figure 3: Scheme of internal and external contacts in eclipse.

For the calculation of the contact points, it is necessary to solve the following system of equations:
$$\begin{cases} x_1 = x + R_2 n_x = A \cdot \sin 2\pi\varphi \\ y_1 = y + R_2 n_y = B \cdot \cos 2\pi\varphi \end{cases}$$

Let the phase of crossing the center of the white dwarf by the limb of the red dwarf that is $\varphi_e$. This phase can be calculated respectively to these formulas for the elliptical and spherical approximations.

In the case, if $r_2 \to 0$,
$$\varphi_e \to (\varphi_{in} + \varphi_{ex})/2$$
$$\sin^2 2\pi\varphi_e = \frac{\frac{b^2}{A^2} - \cos^2 i}{\frac{b^2}{a^2} - \cos^2 i}$$

If $b=a=r_e$:
$$\sin^2 2\pi\varphi_e = \frac{\frac{r_e^2}{A^2} - \cos^2 i}{\sin^2 i}$$

Andronov (1992) got coefficients for determining the radius of the object in the orbital ($a/A$) and polar ($b/A$) plane (that is the plane passing through the center line and the axis of rotation of the system), following the Eggleton's (1983) form. In his work, they are denoted as $\sin\theta(0°)$ ($C=0.4990$, $D=0.5053$) and $\sin\theta(90°)$ ($C=0.4394$, $D=0.5333$), respectively. The elliptic approximation for other angles is correct within 0.2% and 0.5%, respectively. For better accuracy, we used linear interpolation for the ratio of precise/fit values.

We adopted the mass ratio of $q=0.3$ in this system. Then we obtain the following values of the parameters: $r_e/A=0.28103$, $a/A=0.27216$, $b/A=0.26219$.

In Fig. 4, the duration of the eclipse depending on the inclination for both models is shown.

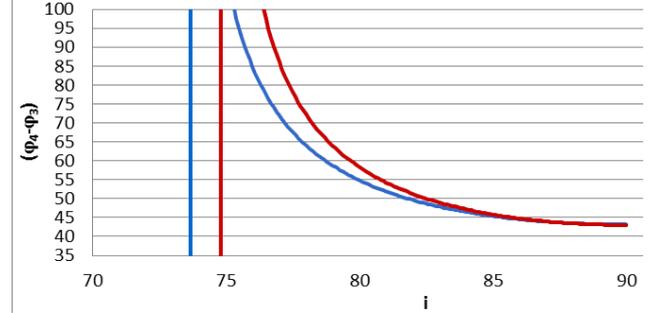

Figure 4: Duration of the ascending/descending branch $\varphi_4-\varphi_3$ (in seconds) as a function of inclination $i$ for the models of circle (blue) and ellipse (red). The vertical lines show limiting values of $i$.

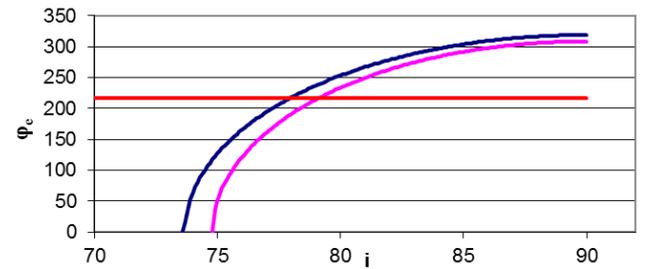

Figure 5: The dependence of the phase $\varphi_e$ of the eclipse of the white dwarf center (expressed in seconds) on inclination for the models of circle (blue) and ellipse (red). The horizontal line shows the observed value.

To determine the duration of the entrance/exit of the eclipse, $\varphi_e$ and the corresponding errors for all parameters, we write the system of conditional equations:

$$T_1 = T_0 + C_1 + P \cdot E - U - V$$
$$T_2 = T_0 + C_1 + P \cdot E - U + V$$
$$T_3 = T_0 + C_1 + P \cdot E + U - V$$
$$T_4 = T_0 + C_1 + P \cdot E + U + V$$

where $T_0$ is an initial epoch,

$C_1$ – correction to the initial epoch,

$P$ – period,

$E$ – number of cycle,

$U$ – the time between crossing the center of the red dwarf limb by the white dwarf and the middle of eclipse,

$V$ – half the length of the ascending / descending branches of the light curve.

From the observations, we have the following values: $T_1$, $T_2$, $T_3$, $T_4$, $E$, $T_0$.

Moments of contacts were obtained from the observations obtained by Dr. Sergey V. Kolesnikov using the 2.6 m telescope named after G.A.Shajn in the CrAO.

Table 1: The moments of eclipse contacts.

| BJD, 2400000+ | Types of contacts | BJD, 2400000+ | Types of contacts |
|---|---|---|---|
| 54946.19335 | 1 | 54949.20429 | 1 |
| 54946.19340 | 2 | 54949.20435 | 2 |
| 54946.19836 | 3 | 54949.20932 | 3 |
| 54946.19842 | 4 | 54949.20937 | 4 |

The parameters were determined in the Excel using the method of the least squares. We have determined the period of the system: $P$=117.18292±0.00014 min.

Parameters such as mass and mass ratio of the stars in this system were taken from the works of other authors, who used not only photometric observations. Details are discussed by Andronov et al. (2014). Considering the phase of the eclipse for white dwarf's center, we obtain two values for inclination of the orbit for the elliptical ($i$=79.1177±0.0075°) and the spherical approximations ($i$=77.1231±0.0029°).

According to the observations, we obtained the duration of the entrance/exit of the eclipse (descending/ascending branches of the light curve): 4.752±0.306 seconds.

We compared this value to that expected for a white dwarf of the accepted mass of $M_1$=0.543$M_\odot$, corresponding to $q$=0.3 and $M_2$=0.163$M_\odot$. From the dependence "mass – radius" for white dwarfs (Andronov and Yavorskiy 1990), we obtained the radius of the white dwarf. Using our calculations for inclination of the orbit, we obtain the duration of the entry/exit (time between external and internal contact) of 63.2 seconds for the elliptical approximation, which is by 13.3 times more than the expected value for the white dwarf.

The Fig. 6 shows two models: the red dwarf, white dwarf at the phase of eclipsing center and the emitting region. Red line shows the elliptical model for the red dwarf, the violet line – circular model. The visible trajectories of the center of the white dwarf for different values of inclination of the orbit are shown. Vertical lines show different values of the parameter $\varphi_e$, including what is observed. From the cross point of the ellipse, one may determine inclination. Also the white dwarf and the emitting area are shown in the same scale.

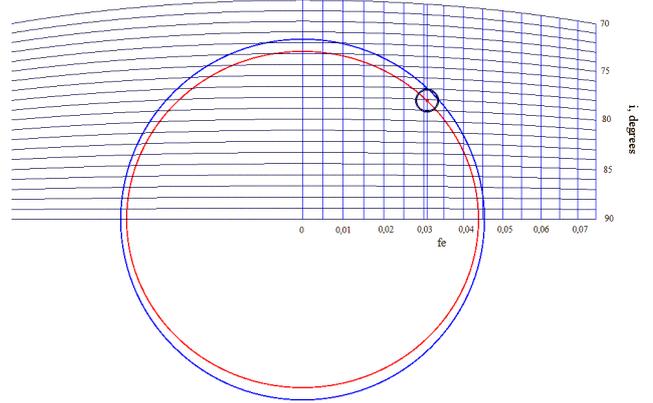

Figure 6: The scheme of eclipse preserving the scale for the numerical values of the model system OTJ 071126 + 440405.

## 4. Conclusions

- The dependencies of the phase of eclipse of the white dwarf's centre and the durations of the ascending and descending branches of the light curve on the binary system's parameters were computed using the spherically – symmetric approximation.
- Similar computations were performed with the more accurate model of the elliptical projection onto the celestial sphere of the companion (red dwarf) that fills its Roche lobe.
- The parameters of eclipses in the classical eclipsing polar OTJ 071126+440405 were estimated.
- The duration of entering/exiting the eclipse is shown to be 13.3 times shorter than the theoretical predictions. Hence, the emitting region is markedly smaller (~1300 km) as compared to the white dwarf's diameter. That is supposed to be the "hot spot" region.

*Acknowledgements.* We would like to sincerely thank Dr. S.V. Kolesnikov for his observations that initiated our research and Dr. V.I. Marsakova for helpful discussions.